# Nuclear quantum effects: their relevance in neutron diffraction studies of liquid water

*Imre Bakó[1], Ádám Madarász[1], László Pusztai[2]*

[1] Research Centre for Natural Sciences, H-1117 Budapest, Magyar tudósok körútja 2., Hungary

[2] Wigner Research Centre for Physics, H-1121 Budapest, Konkoly Thege M. út 29-33., Hungary

**Abstract**

Corrections for nuclear quantum effects (NQE) have been calculated for classical molecular dynamics (MD) simulation models of light ($H_2O$), heavy ($D_2O$) and 'null' $[(H_2O)_{0.64}(D_2O)_{0.36}]$ water. New path integral molecular dynamics (PIMD) simulations have also been conducted for the same systems. NQEs have somewhat smaller influence on the O-D and D-D partial radial distribution functions of heavy water than on the O-H and H-H ones of light water. After correcting for NQEs the O-'H' bondlengths in light and heavy water have become different: the O-D ones are about 0.5 % shorter than the O-H ones. Following NQE corrections, the total RDF of 'null' water does show hydrogen related features at the position of the intramolecular O-'H' peak. Based on a cross-check procedure involving the NQE-corrected total and partial radial distribution functions, it can be stated that concerning the structure of liquid water, the assumption that H and D are equal is valid to a very good approximation for intermolecular correlations. These findings are also supported by our path integral molecular dynamics simulations.

## 1. Introduction

It is well established for molecular liquids that can form hydrogen bonds (HB, H-bond) that their structutural and dynamical properties depend on nuclear quantum effects, too [1-19] Differences in terms of density, dynamic and thermodynamic properties between deuterated and hydrogenated forms of many liquids are well known [20,21], while the nature of deviations in terms of the atomic structure can be quite complex. The main goal of the present work is to shed some light on the structural aspects.

The inclusion of nuclear quantum effects (NQEs), such as zero point energy (ZPE) or tunneling, is possible via path-integral molecular dynamics (PIMD) [22,23] and path integral Monte Carlo (PIMC) [24,25] simulations; unfortunately, routine applications of these methods are prohibitively expensive. Recent results [1,4] revealed that there are at least two different sources of NQE that are relevant from the point of view of the structure: an articulated competition can be seen between the NQEs on the librational modes and those on the stretching modes of hydrogen bonds.

Water is the best known hydrogenated liquid and therefore its structure has been the subject of extensive investigations (see, e.g., [26-30]). Specific to the present study, high energy x-ray, as well as neutron diffraction were applied to measure the temperature variations of hydrogen vs. deuterium isotopic quantum effects on the structure of water [31-38]. The concept that $D_2O$ is a more structured liquid than $H_2O$ at ambient temperature is supported by X-ray [33,35], γ-ray [34], and neutron scattering [31,36] experiments. In other words, heavy water behaves as if it was at a lower temperature, in comparison with light water at the same temperature. The magnitude of the effect depends on temperature: between 279 and 318 K the temperature difference between $D_2O$ and $H_2O$ is found to be in the range of 3.6 to 7.1 K [35,36]. The PIMD estimation of this difference is about 35-40 K, when using simple rigid point charges [13]. On the other hand, more sophisticated flexible anharmonic models [15,16], and recent first principles molecular dynamics studies [36] predict the difference in the range of 5 to 18 K.

A combined neutron and X-ray diffraction study [38] revealed substantial differences in terms of the OH bond length, $d_{OH}$: in $H_2O$ it is longer than the OD bond length, $d_{OD}$, in $D_2O$ by about 3 percent (1.01 Å and 0.98 Å, respectively) Other studies based on oxygen isotopic substitution [36,37] found an about 0.5 % difference between intramolecular O-H and O-D distances (0.990(5) Å and 0.985(5) Å respectively). Results from the latest investigations are in excellent agreement with recent PIMD predictions using an anharmonic flexible force field [15,37].

There are possibilities for calculating NQEs for some thermodynamical properties, like the heat capacity, from classical molecular dynamics (MD) trajectories [39-45]. These methods can estimate quantum corrections by making use of the quantum harmonic oscillator approximation . Recently, some of us proposed a new method [46], called the 'Generalized Smoothed Trajectory Analysis' (GSTA), for the determination of nuclear quantum effects for thermodynamic (heat capacity) and dynamic (diffusion constant) quantities, the dielectric constant and structural properties,  from classical molecular dynamics. A short description of the GSTA method is provided in the Supp. Mat.  The main conclusion was that GSTA is more accurate than the purely classical results, or the 1PT [39] and 2PT [40,41] methods for calculating the heat capacity of liquid water (and about 100 organic liquids). It was also shown that proper application of the GSTA approximation to liquid $H_2O$ significantly decreases the OH intra- (as well as, to some extent, inter-)molecular peak heights.

Neutron diffraction with isotopic substitution (NDIS) [47-51] is an established approach for obtaining detailed information about two-body atomic structure, the partial structure factors (PSF) and partial radial distribution functions (PRDF), in disordered materials. The extraction of PSFs from a series of measurements requires the solution of a set of linear equations, where the measured data (total structure factors) are affected by experimental errors. A rather detailed description of, and possible error propagation through, the data analyses are given in ref. [50]. Most importantly from the point of view of the present work, the analysis of NDIS experiments, with regard to PSFs and PRDFs, is based on the assumption that the structures of the isotopic variants of a material are identical. This assumption is clearly an approximation, particularly in the case of molecular liquids containing the lightest element, hydrogen (and its isotope, deuterium), like liquid water.

As a straightforward extension of the GSTA method, as well as of PIMD simulations, here we calculate the impact of NQEs to $D_2O$ and $(H_2O)_{0.64}(D_2O)_{0.36}$ (so called "null" mixture, see e.g. [47,48,51]), in addition to the case of $H_2O$. Using the GSTA-corrected PRDFs, as well as those obtained from PIMD computations, total radial distribution functions (TRDF) could be calculated, using the coherent scattering lengths of H, D and O.

The calculated total radial functions could then be applied as input for determining the PRDFs with the assumption that the PRDFs of deuterated and hydrogenated samples are identical. PRDFs resulting from this procedure, that mimics the traditional matrix inversion method for evaluating H/D isotopic substitution neutron diffraction data [47,48], can then be compared to PRDFs obtained directly for the isotopic variants from GSTA-corrected MD, as well as from PIMD computations. This way, it is possible to estimate the accuracy of the basic assumption, namely that the structures of the isotopic variants are identical.

As a 'by-product', it is possible to compare GSTA structural results with those of path integral molecular dynamics simulations when the two approaches apply identical force fields and system sizes. These calculations can shed light on the accuracy of the GSTA method since here the PIMD formalism is the relevant reference: it provides *exact* static quantum averages for a chosen interaction potential in the limit of large replica.

## 2. Computational details

### 2.1. Classical molecular dynamics simulations

Dedicated to this investigation, a new set of classical molecular dynamics simulations of liquid water has been conducted, using the SPC/Fw water model [52]. The main reason for taking this potential was that its flexibility made it possible to apply the GSTA correction scheme for vibrational modes . The three systems contained (1) 432 $H_2O$ molecules, (2) 432 $D_2O$ molecules and (3) a mixture of 177 $H_2O$, 56 $D_2O$ and 199 HDO molecules, mimicking the „null" mixture. The edge length of the simulation cell was 2.344 nm in each case, corresponding to a number density of 33.55 $nm^{-3}$. Initial configurations were taken from previous equilibrium simulations of light water at the temperature of 298.15 K [46]. The masses of hydrogen atoms were changed according to the composition of the system. Following a long equilibration procedure (approx. 1 ns), additional 10 ns long production runs were carried out in the NVT ensemble, during which particle configurations were saved after every 40000 steps (20 ps) for statistical analyses (500 configurations). Other simulation settings were taken from our previous work [46]. Each trajectory from the saved configurations (every 100ps) was run for 10 ps at constant *NVT,* followed by 10 ps of *NVE* simulation. GSTA filtering was performed over the last 243 fs long trajectories, resulting one quantum corrected configuration. PRDFs were determined from 500 independent, quantum corrected configurations.

In addition, we repeated the above procedure for the qSPC/Fw [15] force field, which is a specifically designed modification of the SPC/Fw potential, in order to determine the NQE using the path integral molecular dynamics formalism.

*2.2. Path integral molecular dynamics simulations*

Settings for PIMD simulations were taken from the literature [15]. Here, PIMD simulations were carried out at 298.15 K, by applying the Nose-Hover chain thermostat. The three simulatyed systems consisted of (1) 423 $D_2O$, (2) 423 $H_2O$, and (3) a mixture of 59 $D_2O$, 177 $H_2O$ and 199 $D_2O$ molecules („null mixture") in cubic boxes, at the same density as we applied in the „classical" simulations. The time step was 0.1 fs. The main differences from the literature calculation are that we used 32 beads instead of 24, and 432 water molecules instead of 216. After a 1 ns long equilibration, particle configurations were collected from another 1 ns long trajectory, using snapshots taken at every 1 ps: that is, the radial distribution functions were determined from 32000 configurations. The computational time for this calculation was about one week on 32 CPU-s on our home cluster.

*2.3. The total radial distribution function formalism for neutron diffraction*

The total radial distribution function, that represents a measurable description of the two-particle level structure of multi-component liquids, may be calculated from the partial rdf's according to the equation

$$G(r) = \frac{\sum_{\alpha>\beta}\sum(2-\delta_{\alpha\beta})b_\alpha b_\beta g_{\alpha\beta}(r)}{\left(\sum_\alpha b_\alpha\right)^2} = \sum_{\alpha>\beta}\sum w_{\alpha\beta} g_{\alpha\beta}8r) \qquad (1)$$

where $g_{\alpha\beta}(r)$ are the partial radial distribution functions (PRDF), $b_\alpha$ are the neutron scattering lengths, and $w_{\alpha\beta}$ are the neutron scattering weighting factors. This equation is valid if the coherent scattering length does not depend on the scattering variable. The coherent scattering lengths for H, D and O are -0.373, 0.667 and 0.583 fm, respectively. The isotopic substitution technique, based on the –assumed— equivalence of the OH and OD, as well as the HH and DD PRDFs, can be applied for liquid water using the total radial distribution functions of $D_2O$ and $(H_2O)_{0.64}(D_2O)_{0.36}$ (so called "null" mixture, see e.g. [47,48,51]), in addition to that of $H_2O$. Matrix formalism can be used to describe the relation between experimental total and partial radial distribution functions

$$F = W * y \qquad (2)$$

where a column vector **y** contains the partial radial distribution functions (*prdf*) at a given $r$ distance, **W** is the neutron scattering weighting matrix. The row vectors of matrix **F** are the three independent experimental total radial distribution functions, as defined by Eq. (1), weighted according to the relevant neutron scattering lengths. Equation (2) of the isotopic substitution method can be solved for column vector **y** by using a matrix inversion technique.

## 3. Results and discussion

### *3.1. Partial radial distribution functions*

In our earlier work [46] we showed that a good agreement between calculated and experimental values of the heat capacity of water can be reached by using the SPC/Fw water interaction potential [15]. Here we wish to benchmark the GSTA filtering method against the full PIMD calculation, in terms of the partial and weighted total radial distribution functions of liquid water at room temperature. Fig. 1 compares the PIMD and the NQE corrected "classical" PRDFs of liquid $H_2O$. It can be concluded that in the range of intramolecular distances, the two approaches agree very well.

It is also instructive to compare the bond lengths $d_{OH}$ and $d_{OD}$, whether the differences claimed to have been observed experimentally [36-38] are reproduced after the GSTA treatment. Characteristic intramolecular OH and OD distances are presented in Table 1. In the classical simulations with the SPC/Fw (qSPC/Fw) model [52,15], each O-'H' intramolecular distance was 1.031(1.019) Å. As a result of the corrections for NQEs, $d_{OH}$ and $d_{OD}$ have become 1.042 and 1.037 nm (1.0328 and 1.026 Å), respectively, as a result of GSTA filtering. The mean square deviations of the distances mentioned above in both methods applied are in the range of 0.05 and 0.07 Å, and the uncertainties of the average values are in the range of 0.001 and 0.002 Å. That is, the classical O-'H' bond length has increased for both light and heavy water, and the bond length in light water is longer by about 1 %.

On the other hand, OH and OD intramolecular distances calculated from PIMD simulations are shortening very slightly when using the SPC/Fw and qSPC/Fw force fields. In each case, using the Student T-test, we found that differences between classical and PIMD values, in terms of distances and angles, were significant only at a p = 0.05 level. That is, the difference between OH and OD distances is negligible, as well as the difference between classical and PIMD values. This disagrees somewhat with results from GSTA filtering.

As a summary concerning effects of NQEs on the first, intramolecular, maxima of the OH, OD, HH and HD PRDFs, broadening of the 'classical' peaks is always present and the peak heights always decrease in the order of classical > OD (DD) > OH (HH), for both the PIMD and GSTA-corrected cases. On the other hand, the validity of the tiny shifts of the positions of the maxima cannot be confirmed yet: one can say that the GSTA-corrected values follow experimental findings more closely (but these experimental findings do not agree very well).

At greater distances, where anharmonic movements play a larger role (e.g., beyond the first maximum of the OO PRDF), small but well-defined deviations are observed. For both OH/OD and HH/HD partials, differences between classical and PI MD are visibly larger than those between classical and GSTA-corrected classical MD. The NQE effects calculated from PIMD simulation cause a larger shift (appr. 0.03 Å) of the intermolecular OO distances than those from the GSTA method. In addition, the application of the exact NQE formalism (PIMD) caused a larger effect on the height of the first peak of the O-O partial radial distribution function. Note that all of these deviations are still well within the estimated experimental uncertainties.

Figure 2 shows comparisons between classical and NQE corrected PRDFs for the three isotopic variants of liquid water. There are a few things to note here: (i) the NQE effect is largest for the intramolecular distances, and it is larger for the O-`H` bonding distance; (ii) intermolecular distances are much less affected; (2) the NQE effect is somewhat larger for H-containing pairs than it is for D-containing pairs; (3) the NQE effect is visible on the O-O PRDF, as well. It may be stated that that the influence of NQEs on heavy and "null" water is qualitatively similar to what has been found for pure $H_2O$ [46]: quantum effects cause visible widening the sharp features characteristic to the purely classical results, particularly at low *r*.

Observations described above are also true for comparisons of PIMD and classical simulation results (see Fig. 1). In this particular setup, the effect of NQE on the OO pair correlation function is more pronounced in PIMD calculations. This latter is a rather spurious finding that goes against commonsense expectations: lighter elements and shorter distances must be affected to a larger extent than heavier atoms and longer distances. The reason behind this odd behavior is thought to be the potential model applied.

We now turn to a detailed investigation of the NQE-corrected PRDFs. First, Fig. 2 compares some parts of the PRDFs calculated directly from the classical trajectories: clearly, there are no visible differences between functions corresponding to the different isotopic compositions. This situation changes when GSTA corrections for the NQEs are applied (see Figs. 1 and 2), particularly for the intramolecular O-'H' distance: the OD maximum is more than 20 % higher than the OH one. As far as the first intermolecular O-'H' maximum (corresponding to the hydrogen bond) is concerned, the difference is much smaller: the OD peak is only about 5 % more intense than the OH one. Similar effects, although on a smaller scale, can be observed on the 'H'-'H' PRDFs. Results mentioned above for qSPC/Fw simulations are presented in Figs. S1-S3 of the Supp. Mat. The conclusions are the same as for SPC/Fw simulations.

We have also applied the GSTA method to well equilibrated $D_2O$ and $H_2O$ systems (approx. 100000 configurations) obtained from published ab initio molecular dynamics simulations of 128 water molecules using the BLYP-D3 density functional [53]. Computational details of this simulation can be found in the Supp. Mat. GSTA filtering was performed and the PRDFs were determined from 500 filtered configurations. Our studies show that the calculated PRDFs do not change significantly when the number of filtered configurations is varied between 100 and 1000. Calculated $d_{OH}$ and $d_{OD}$ for the NQE corrected AIMD [17] trajectories were found to be 1.001(9) (MSD: 0.05) and 0.997(0) (MSD: 0.04) Å , respectively. In contrast, both the OH and OD distances obtained directly from AIMD simulations (without taking the NQE effects into account) is about 0.990 (MSD: 0.02) Å. The calculated NQE corrected PRDF are shown in the Supp. Mat., Figs S4-S7. The difference between $d_{OH}$ and $d_{OD}$ is about 0.4 % here. Machida et al. reported $d_{OH}$ and $d_{OD}$ distances in liquid $H_2O$ and $D_2O$ using PIMD/AIMD simulation: they were found to be 1.014 and 1.008 Å, respectively, which is a difference of about 0.6 % [11].

In summary, the difference between $d_{OH}$ and $d_{OD}$ was found to be between 0.4 and 1 % for the three, rather different, approaches (GSTA-corrected classical MD; GSTA-corrected ab initio MD; PIMD/AIMD [11]) mentioned above, which is a rather good agreement – and these values are also consistent with experimental values obtained on the basis of NDIS experiments [36,37]. On the other hand, PIMD with the two classical potential functions considered here offers a slightly, but qualitatively different outcome, the very small shortening of the bond lengths. It may then be conjectured that the SPC/Fw and qSPC/Fw water potentials are perhaps not the best choices for representing the structure of 'real' liquid water.

*3.2. Total weighted radial distribution functions*

Next, total radial distribution functions have been composed from the GSTA filtered PRDFs, for each of the three isotopic variations of simulated water, by using coherent scattering lengths for H, D and O as -0.373 0.667 and 0.583 fm, respectively. Resulting TRDFs are displayed in Figs. 3 and 4. Concerning pure light and heavy water (Figures 4 a and b), the differences between classical and NQE corrected totals, up to about 0.2 nm, reflect what we have already seen for the partials, i.e. quantum corrections lead to substantial dampening and widening of maxima. The TRDF of light water exhibits a large negative extremum at around 0.1 nm: this is completely normal, since the coherent neutron scattering length of H is negative.

However, a more detailed explanation is necessary concerning the TRDF of the "null" mixture, whose composition (64 % light water and 36 % heavy water) was set in order to obtain an average coherent scattering length of zero for hydrogen (0.64x(-0.37) + 0.36x0.67 = ca. 0). This way, in principal, the TRDF of the 'null' mixture should be identical to the O-O PRDF, as all contributions with hydrogen are zero (due to the zero average scattering length of hydrogen). The classical TRDF of 'null' water does follow this 'recipe' and therefore, the strong extrema found for light and heavy water at around 0.1 nm disappear (with a very good approximation). The difference between data from PIMD and GSTA calculations mentioned earlier is clearly visible in Fig 4c.

Note that from a simulation trajectory, where the coordinates of each atom are available, it is possible to calculate the TRDF not only by taking an average coherent scattering length for hydrogens, but also, to determine each PRDF exactly, by assigning the coherent scattering lengths of H and D according to the actual identity of a given 'H' atom. This way, there are six

PRDFs in 'null' water: HH, HD, HO, DD, DO and OO. When one calculates the TRDF of GSTA filtered null water then a rather unexpected curve emerges (see Figure 4 c): there is well developed maximum at around 0.1 nm. This is due to the fact that quantum corrections (such as GSTA for taking NQEs into account) influence OH and OD (and HH, HD and DD) differently.

We stress that if instead of the direct calculation described just above, we would count every 'H' atom with the average coherent scattering length, that is zero to a very good approximation, then the extra intensity at around 0.1 nm would still not appear. However, when one executes a real neutron diffraction experiment on real 'null' water then we should be able to detect the maximum corresponding to the intramolecular O-'H' distance – in real 'null' water, none of the 'H' atoms are 'average'. That is, a very carefully executed neutron diffraction measurement on 'null' water would, in principle, be capable of providing information on NQEs, through the maximum that would appear around 0.1 nm. Unfortunately, the exceptionally high level of incoherent scattering from H (see, e.g. [53]), in combination with the problem of not good enough statistics at the necessary high momentum transfer values makes a definitive measurement and subsequent corrections virtually impossible as yet.

If the comparison is performed at the level of the total radial distribution functions then the two methods show a good agreement with each other, and with the experimental data, as well. This compassion is shown in Fig 4. Results for qSPC/FW simulations are presented in the Supp. Mat., Figs. S1 and S2.

### 3.3. On the validity of H/D isotopic substitution in neutron diffraction

The capability of determining quantum corrections for H and D (and O, but this latter is unimportant from the point of view of H/D isotopic substitution) provides us with a unique possibility: we have now everything at hand for assessing whether the frequently used approximation of the equality of H and D while determining PRDFs from NDIS experiments is valid. For this, we have used the TRDFs shown in Figure 4 for separating the partials via the traditional matrix inversion method [47-51].

In Figure 5, comparisons of three OH and HH PRDFs are shown: (1) separated with the assumption that H=D, (2) OH in $H_2O$, as calculated from the classical MD trajectory and corrected by GSTA, and (3) OD in $D_2O$, as calculated from the classical MD trajectory and corrected by GSTA. Also in Figure 6, comparisons with PIMD results are provided. As it could be expected, there are differences between these functions, particularly in the intramolecular region: maxima for case (1) are higher by about 10 % than those for pure light water. Overall, the H=D case seems to be somewhat closer to the case of pure $D_2O$. On the other hand, beyond the intramolecular region, i.e. where it is really important from most points of view, the H=D assumption works really well: the curves run very nearly together (particularly with those of pure heavy water). This statement is also supported by PIMD results (Fig. 5). It must be stressed that, unfortunately, systematic errors associated with NDIS experiments for water are almost certainly larger than deviations detected in the intermolecular regions in Figure 5.

## 4. Conclusions

In summary, new classical molecular dynamics simulations, as well as new PIMD computations, for light ($H_2O$), heavy ($D_2O$) and 'null' ($H_{0.64}D_{0.36}O$) water have been carried out. Corrections for nuclear quantum effects have been calculated for the PRDFs by using the recently developed GSTA method [46] that has been shown to be superior to earlier, similar protocols. Concerning the structure of liquid water, the following conclusions may be drawn:

(i) As it may have been expected, NQEs have somewhat smaller influence on the 'O-H' and 'H-H' PRDFs of heavy water than on those of light water.

(ii) As a result of GSTA corrections, the O-'H' bond lengths in light and heavy water have become different, the O-D ones in the latter being about 1 % shorter. This is consistent with findings of earlier experimental investigations [36-38], as well as with PIMD/AIMD results [11].

(iii) As a result of different NQE corrections for light and heavy water, the total RDF of 'null' water does show hydrogen related features, most noticeably, at the position of the intramolecular O-H peak.

(iv) Based on the cross-check procedure described above (calculate GSTA-corrected PRDFs for light, heavy and 'null' water; compose total RDFs; separate partials by assuming that H=D), it can be stated that concerning the structure of liquid water, the assumption that H and D are equal is valid to a very good approximation for intermolecular correlations. This statement is valid for our new PIMD results (with SPC/Fw and qSPC/Fw potential functions), too.

(v) Reasonably good agreement between GSTA filtered and PIMD simulation results has been found, particularly in the region of intramolecular distances.

**Ackowledgements**

This work was supported by the National Research, Development and Innovation Office of Hungary (NKFIH, Grants No. K 124885 and KH 130425). The authors would like to thank Ari P. Seitsonen for fruitful discussions and for the AIMD configuration sets for $D_2O$ and $H_2O$.

**Table 1** Characteristic OH and OD intra and intermolecular distances (Å) and HOH intramolecular angles from classical, GSTA filtered and PIMD simulations using SPC/Fw and qSPC/Fw force fields. (Bond lengths are given in Å.) Statistical uncertainties of the intramolecular distances are in the range of 0.001 and 0.002 Å, those of intermolecular distances are in the range of 0.005 and 0.007 Å, whereas those of the bond angle is in the range between 0.2 and 0.3 degrees.

|  | SPC/Fw | qSPC/Fw |
|---|---|---|
| „classical" | 1.0312 | 1.0199 |
| PIMD OH | 1.0302 | 1.0190 |
| PIMD OD | 1.0306 | 1.0194 |
| GSTA OH | 1.0425 | 1.0312 |
| GSTA OD | 1.0373 | 1.0264 |
| HOH angle |  |  |
| GSTA | 107.45 | 105.88 |
| PIMD | 107.78 | 106.28 |
| „classical" | 107.70 | 106.10 |
| H-bonded O..H distances |  |  |
| GSTA OH,OD | 1.8411, 1.8420 | 1.8443,1.8445 |
| PIMD OH ,OD | 1.8800, 1,8651 | 1.8812,1.8739 |
| classical OH | 1.8410 | 1.8449 |

**Figure 1** The original („classical"), GSTA filtered classical, and PIMD partial radial distribution functions for $H_2O/D_2O$, using the SPC/Fw force field for water.

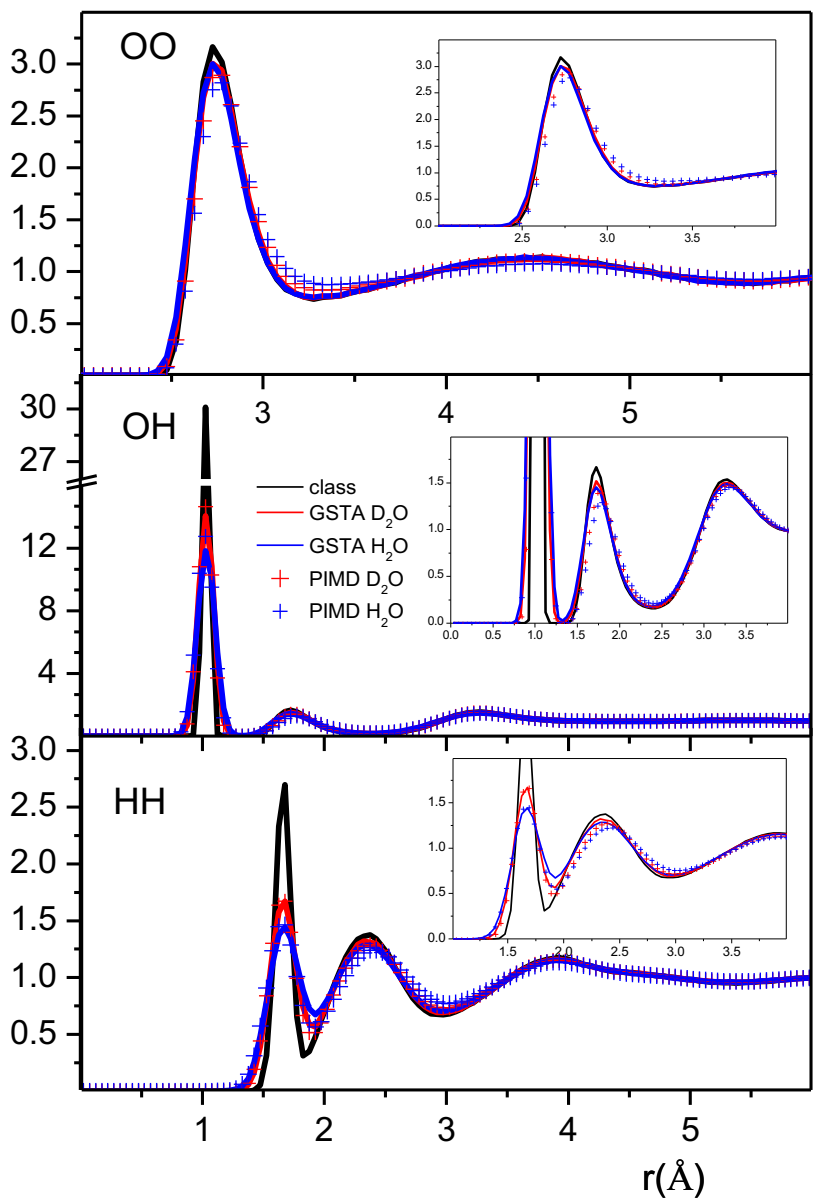

**Figure 2** The original and GSTA filtered partial radial distribution functions in $D_2O$, $H_2O$ and $O_2O$.

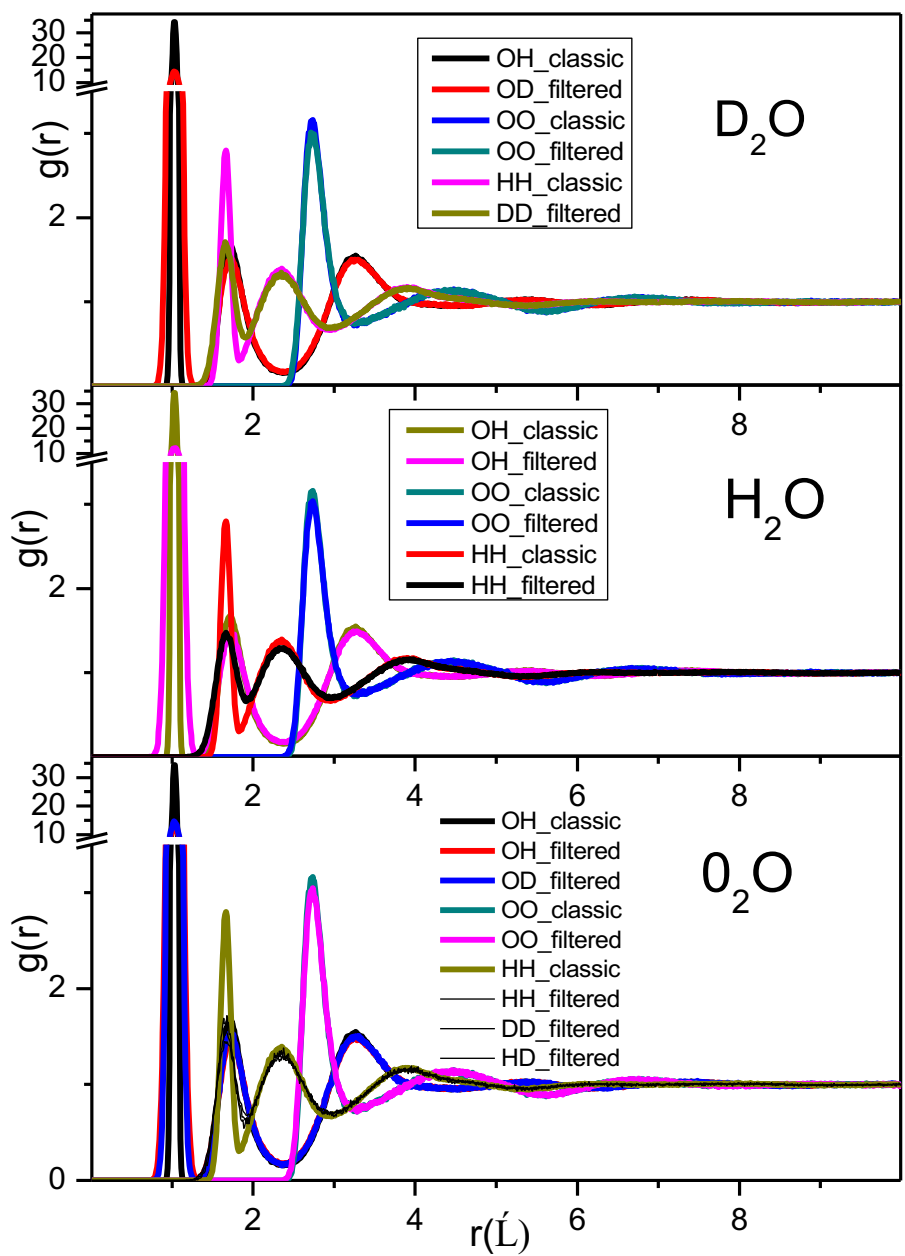

**Figure 3** Total radial distribution functions of $D_2O$, $H_2O$ $0_2O$ from GSTA filtered classical MD, PIMD and classical MD simulations, using the SPC/Fw water potential.

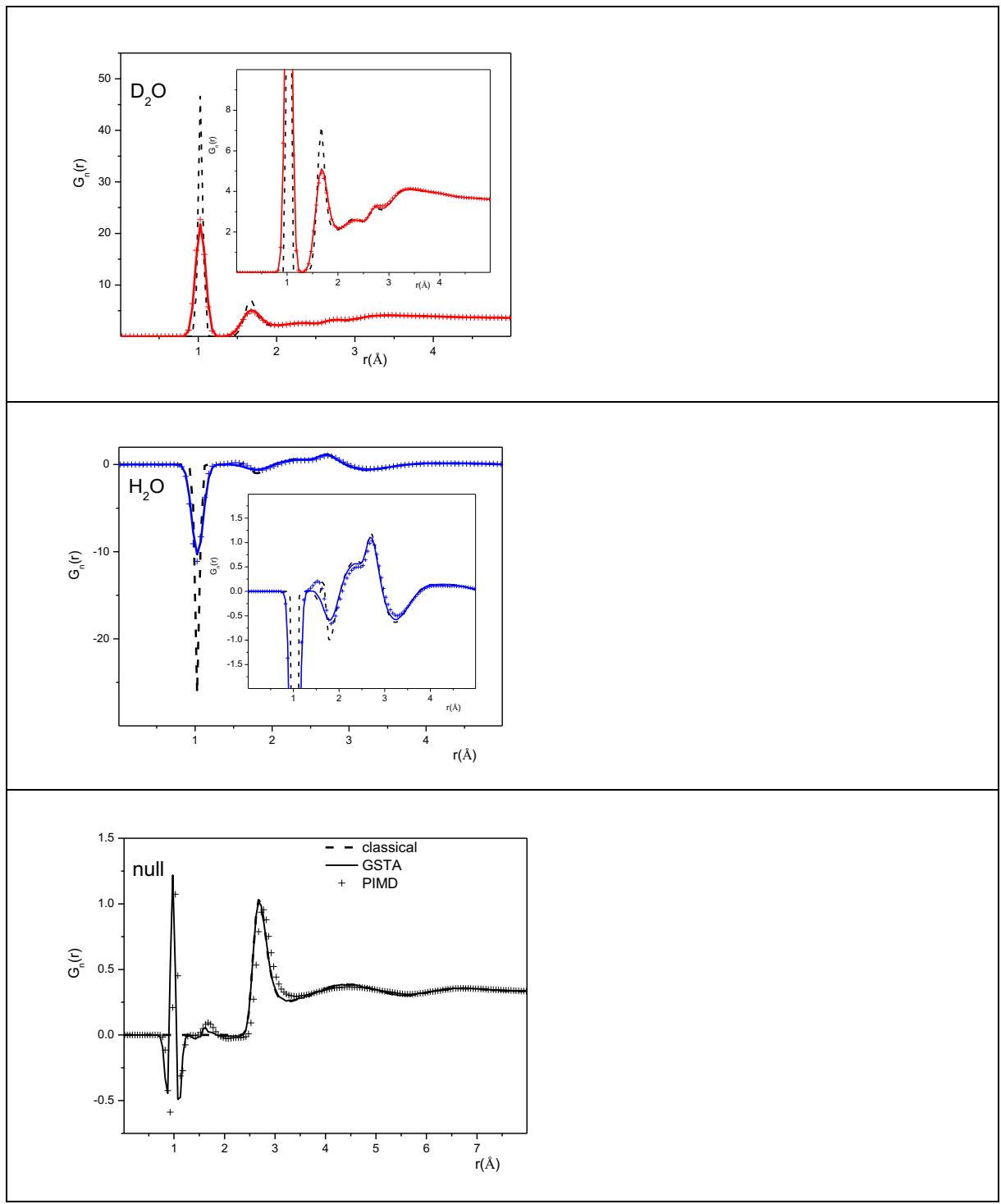

**Figure 4** Total radial distribution functions of $D_2O$ and $H_2O$ from GSTA filtered classical and PIMD SPC/Fw simulations, in comparison with experimental data [47].

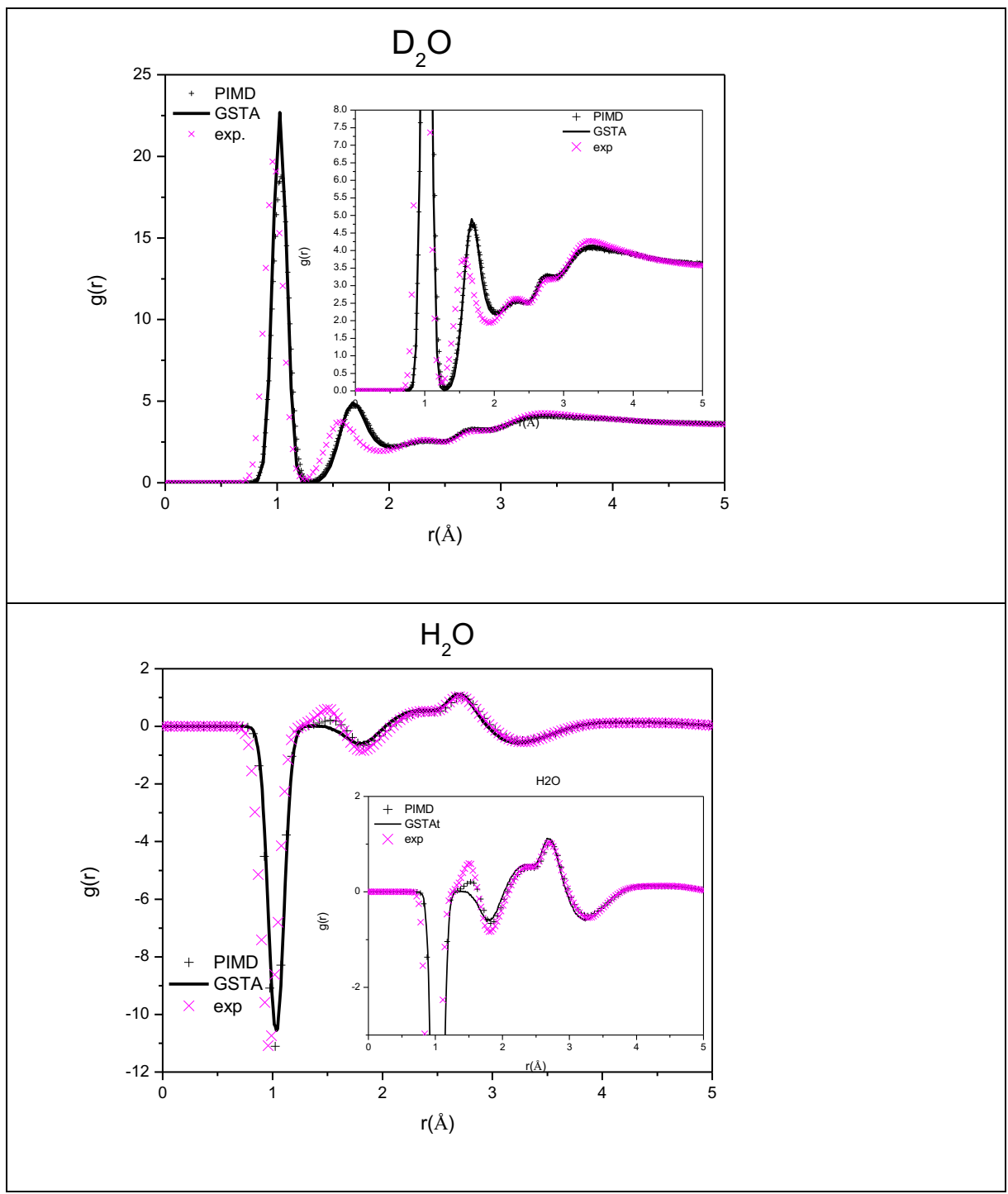

**Figure 5 a.b** Comparisons of three OH (parts a) and c)) and HH (parts b) and d)) PRDFs are shown: (1) separated from the total RDFs (see Fig. 3) with the assumption that H=D, (2) OH and HH in $H_2O$, as calculated from the MD trajectory and corrected by GSTA, and (3) OD and DD in $D_2O$, as calculated from the MD trajectory and corrected by GSTA (parts a) and b). Parts c) and d): same as above, but as calculated from the PIMD trajectory. In both simulations the SPC/Fw water model was used. The insets in parts a) and c) show the maxima of the intramolecular peaks of the OH/OD PRDF.

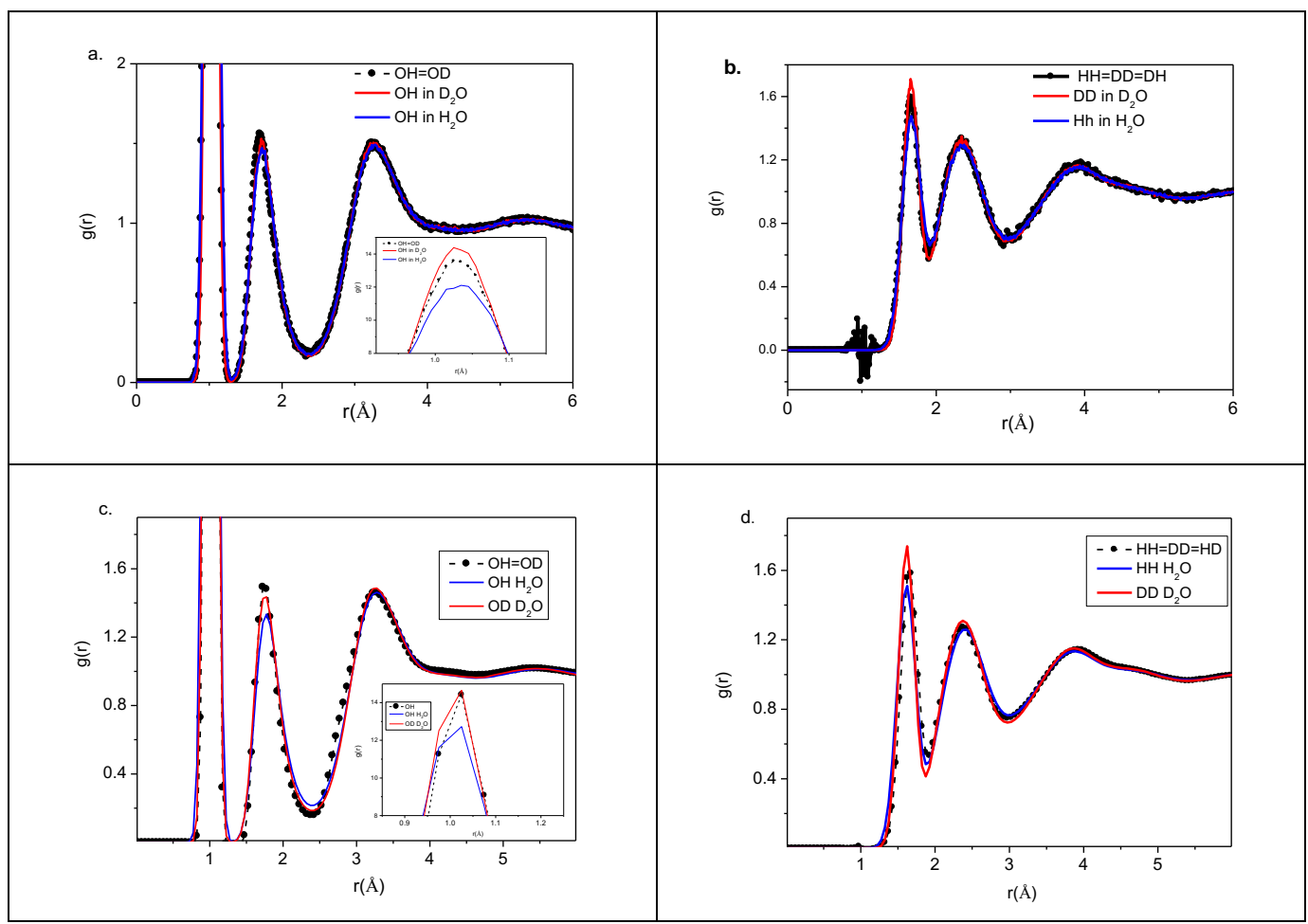